\documentclass[rnote]{aa}
\usepackage{natbib,graphicx}
\usepackage{rotating}
\usepackage{amssymb, amsmath}
\usepackage[varg]{txfonts}
\usepackage{color}
\usepackage{journals}
\begin{document}

\title{The nature of the nearest compact group of galaxies\\ from precise
  distance measurements}

\author{Gary A. Mamon}
\institute{Institut d'Astrophysique de Paris (UMR 7095: CNRS \& UPMC), 98 bis
  Bd Arago, F--75014 Paris, France, {\tt gam AT iap.fr}}
\date{Received 21 March 2008 / Accepted 18 April 2008}
\authorrunning{Mamon}
\titlerunning{The nature of the nearest compact group of galaxies}

\abstract
{Compact groups (CGs) of galaxies, similar to those catalogued by Hickson, appear to be
the
  densest galaxy structures in the Universe. Redshift information is 
  insufficient to determine whether a CG is roughly as dense in three
  dimensions as it appears in projection, or whether it is caused by a chance
  alignment along the line of sight within a larger galaxy system.}
{Recent precise distance measurements help probe the nature of the nearest
  CG, situated in the Virgo cluster, whose dominant member is
  M60. } 
{The isolated status of the CG is reassessed with recent photometry and a
  statistical analysis is performed on the surface brightness fluctuation (SBF)
  distances measured by Mei et al. in Virgo, for 4 of the 5 CG
  members. } 
{The neighboring galaxy
  NGC~4606 appears (with 80-90\% confidence)
to be too faint to affect the isolated status of
  the CG.
Taken at face value, the SBF distances suggest that M59 and NGC~4660 lie roughly 2 Mpc
  to the foreground of M60, NGC~4638, and the bulk of the
  Virgo cluster. 
The statistical analysis confirms that
the CG is, indeed, the result of a chance alignment of its galaxies, with
NGC~4638 lying at least 800 kpc further away (with 99\% confidence)
than either 
  M59 or NGC~4660. 
 The first two galaxy distances are consistent with M59 and
  NGC 4660 constituting a tight pair. 
The dominant galaxy, M60, is at least 440 kpc more distant (95\%
  confidence) than the M59+NGC~4660 pair, and over 1 Mpc (99\%
  confidence) more distant if one uses the broken linear calibration
  of the SBF distances. }
{This work constitutes the first direct analysis of the nature of a compact
  group of galaxies.
Chance alignments of galaxies represent 
a realistic alternative to truly dense groups to explain 
the nature of CGs. 
With current SBF distance accuracies, one could determine the nature of HCG~68 in the same
  way.} 


\maketitle

\section{Introduction}
\label{intro}
Compact groups of galaxies (CGs) appear to be the densest known
multiple galaxy systems 
(with mean densities $\sim 10^4$ times the critical density of the
Universe). The CG catalog by far the most studied is
the one assembled by \cite{Hickson82}, who 
visually searched  the
POSS~I photographic plates  for isolated
groups of at least 4 members  within 3 magnitudes of the brightest, whose
mean surface brightness exceeded a given threshold. The mean surface
brightness is measured within the smallest circumscribed circle (hereafter,
\emph{Hickson circle})
containing
the centers of the galaxies. The isolation criterion specifies that there are
no galaxies within 3 mag (in the $R$ band) from the brightest CG member
within an 
\emph{isolation ring} extending from the Hickson circle to
a concentric circle 3 times wider.

A spectroscopic followup by \cite{HMdOHP92} revealed that among the 100 Hickson
compact groups (HCGs), only 69
groups had at least four members with accordant velocities (within $1000 \,
\rm km \, s^{-1}$ from the median).
Still, it is unclear whether these 69 HCGs are roughly as dense in
three dimensions as
they appear to be in projection \citep{HR88}, 
or whether they are caused by chance
alignments of galaxies along the line of sight (\citealp{Rose77} for the
elongated CGs; \citealp{Mamon86} and \citealp{WM89} for most HCGs). The
galaxies in a
chance alignment lie in a looser group \citep{Mamon86,WM89}, a cluster
(\citeauthor{WM89}), 
or an even longer cosmological filament \citep*{HKW95}.
Although HCG galaxies display numerous signs of dynamical
interaction with close neighbors (\citealp{Hickson97rev}, and references
therein), those HCGs caused by chance alignments 
are expected to be binary-rich
\citep{Mamon90_IAUC124_CAs,Mamon92_DAEC}, and these binaries should 
explain --- to first order
---
the frequency of interacting galaxies \citep{Mamon92_DAEC}.

Motivated by \citeauthor{WM89}'s prediction that the frequency of chance
alignments increases with the number of galaxies in the parent system,
I had searched the Virgo cluster for CGs meeting
Hickson's selection criterion, and indeed found a CG, composed of
\object{M60}, \object{M59}, \object{NGC~4660}, \object{NGC~4638}, and
\object{NGC~4647} \citep{Mamon89}. 
\begin{figure}[ht]
\centering
\includegraphics[width=0.7\hsize, angle=-90, bb=100 224 492 618, clip=true]{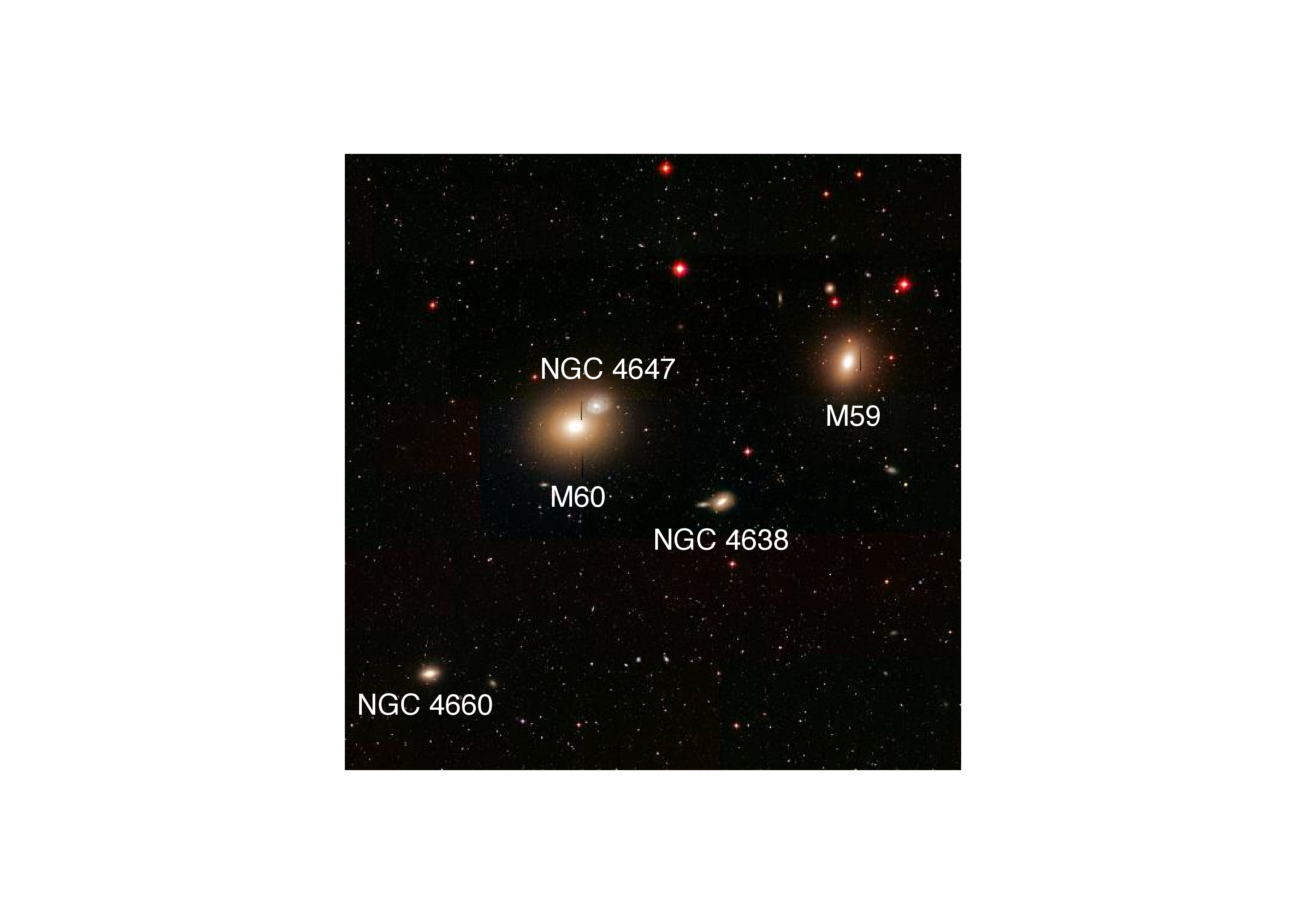}
\caption{SDSS mosaic of the M60 compact 
group in RGB display using the $g$, $r$ and
  $i$ SDSS images. The image is $54\farcm1$ wide.
The \emph{dark vertical lines} in the Northern part of M60 and the Western
part of M59 are image artefacts.}
\label{SDSS}
\end{figure}
Figure~\ref{SDSS} displays a view of this compact group (hereafter called
the M60 CG), taken from the Sloan Digital Sky Survey (SDSS).

The M60 CG had been missed by \cite{Hickson82}, because
\object{NGC~4606}, an  Sa galaxy
lying at 1.98 Hickson circle radii from the group center,
was only 2.4 mag fainter than M60 in the $B$ band. Even after
a crude
extrapolation to $R$ magnitudes for the different morphological types of the two
galaxies, NGC~4606 was still slightly less than 3 mag fainter than M60,
so
NGC 4606
caused the CG to fail Hickson's isolation criterion. 
When I discovered this CG \citep{Mamon89}, I noticed that more
accurate $B$-band photometry indicated that NGC~4606 was 2.88 mag
fainter in $B$ than M60, which, after the crude correction for 
morphological types, suggested that NGC~4606 was at least 3 mag fainter
than M60 in the $R$ band. I therefore argued that NGC~4606 did not affect
the isolated status of the CG, and concluded that the M60 CG
was the nearest HCG-like group.

The nature of CGs within clusters is by no means clear. While
chance alignments are expected to be frequent in clusters \citep{WM89}, one
also expects to see
groups falling in or bouncing out of clusters before becoming 
dynamically mixed with
their host cluster. The tidal field of the
cluster should truncate the infalling groups after their first passage,
leaving 
groups with high density close to that of the cluster at pericenter
\citep{Mamon95_Chalonge,GHO99}, which is of the same order as the mean
density of the M60 CG.

Recent measurements by \cite{Mei+07} of the distances to  Virgo
  elliptical and   lenticular galaxies, through the accurate surface
  brightness fluctuation (SBF)
  method \citep{TS88}, permit to check whether the M60 CG is dense in 3D or whether
  it is caused by a chance alignment of galaxies.
This is the first CG meeting the HCG criteria that is sufficiently
nearby to have its nature determined by SBF distance measurements.

In this \emph{Research Note}, 
I first re-investigate in Sect.~\ref{isol} whether the
latest photometric measurements confirm that NGC~4606
is too faint to be considered a contaminant of the isolation ring. 
I then present briefly, in Sect.~\ref{data}, the SBF distance
measurements. 
In
Sect.~\ref{anal}, 
I estimate lower limit for the line-of-sight separations of the M60 CG galaxies, 
given the SBF distance measurements and their errors, to decide whether 
the M60 CG
is a chance alignment of galaxies or a true dense group.
Finally, I discuss in Sect.~\ref{discus} what should be the maximum line-of-sight
  size of a 
  dense group, and compare with the lower limits determined in
  Sect.~\ref{anal}. I also investigate which other HCGs are both close
  enough and with sufficient numbers of bright early-type galaxies to have
  their nature determined by SBF
  measurements with present-day accuracies.

\section{Is NGC 4606 sufficiently bright to affect the isolated status of the M60
  compact group?}
\label{isol}

\begin{figure}[ht]
\centering
\includegraphics[width=0.8\hsize, angle=-90, bb=85 305 365 584, clip=true]{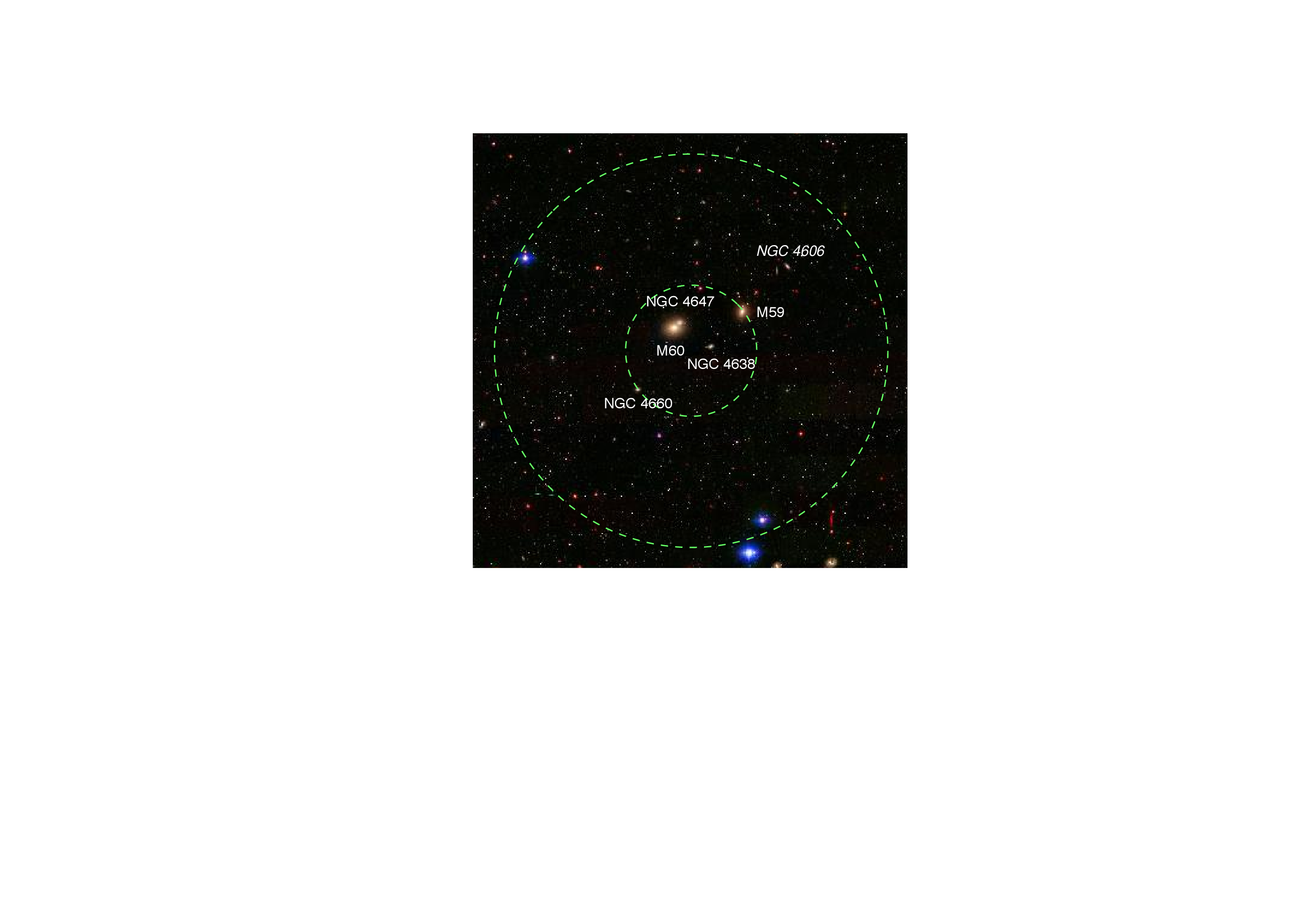}
\caption{SDSS mosaic of the M60 compact 
group and its environment in RGB display
  using the $g$, $r$ and 
  $i$ SDSS images. The image is $2\fdg7$ wide. The \emph{inner} and
  \emph{outer circles} show the limit of the group and the outer radius of
  the isolation ring, respectively. The two bright blue objects in the ring
  are foreground stars. }
\label{SDSSwide}
\end{figure}

Figure~\ref{SDSSwide} shows the large-scale environment of the M60 CG, with
NGC~4606 lying within the isolation ring.
Is NGC~4606 bright enough to prevent the M60
  CG from being isolated? Following Hickson's original isolation criterion, 
the M60 CG is isolated if
NGC~4606 is over 3 magnitudes fainter than M60 in the $R$ band.

Unfortunately, there is yet no good $R$-band photometry for
NGC~4606 and M60. The 6th Data Release of the Sloan Digital Sky Survey (SDSS)
obtained photometric measurements for NGC~4606, but with bad photometric 
flags.
As for measurements of other bright galaxies, the SDSS photometric
measurement for M60 is off by $\approx 3$ magnitudes (Mamon et al., in
prep.), probably 
because of poor background subtraction. 
For these reasons, neither galaxy has SDSS photometry in the NASA/IPAC
Extragalactic Database (NED).

I attempted to measure the photometry of these two galaxies directly from
  the SDSS images.
A SExtractor \citep{BA96} extraction of NGC~4606 (using large
$512\times512$ pixel tiles to estimate the background, thus avoiding
an overestimate of the 
background at the position of the large galaxy), gave
$r=11.77\pm0.00$ (while its 
magnitude in the 
SDSS database is $r=12.22\pm0.00$).
On the other hand,
M60 is located near the edge of its $14'x10'$ scan, and its image
almost fills the entire scan, so that 
the background subtraction is uncertain, which leads to important
uncertainties in the
photometry for M60 (one can also distinguish different background levels in the
SDSS mosaic of Fig.~\ref{SDSS}).


\begin{figure}[ht]
\centering
\includegraphics[width=0.8\hsize]{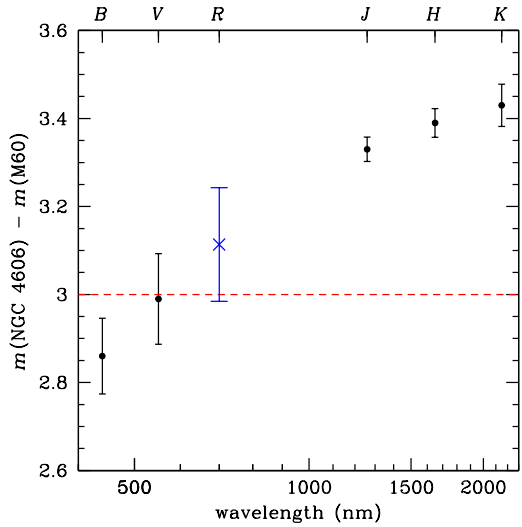}
\caption{Magnitude difference between NGC~4606 and M60 in different
  wavebands: $B_T$, $V_T$ from the RC3 \citep{RC3}  
and $J$, $H$, $K$ from 2MASS \citep{Jarrett+00}, all read in
  NED. The \emph{cross} is the spline fitted value for the $R$ band (with
  Monte-Carlo $1\,\sigma$ error).} 
\label{dmags}
\end{figure}
Figure~\ref{dmags} shows the difference in magnitude between
NGC~4606 and the giant elliptical M60
for different wavelengths. Given its morphological type, NGC~4606 is bluer
than M60 (as 
can be seen in Fig.~\ref{SDSSwide}).
Fitting a cubic spline to the magnitude difference as a function of log
wavelength, NGC~4606
is found to be
3.11 mag fainter than M60 in the $R$ band. 
However, assuming Gaussian-distributed magnitude errors, a simple Monte
Carlo
analysis  (with $10^5$ trials) shows that only 81\% of the time is the magnitude
difference in the $R$ band greater than 3 magnitudes.

Alternatively, the total $R$-band photometry of M60 and
NGC~4606 can be found by extrapolating the $B$ or $V$ total photometry from
the RC3 \citep{RC3} using
$B-R$ or $V-R$ colors measured in annuli at roughly half the luminosity.
M60 has $B_T=9.81\pm0.05$ \citep{RC3} and a color 
$B-R=1.59$ at the effective radius (from 
\citealp{Peletier+90}), yielding $R_T\simeq8.22\pm0.05$.
NGC~4606 has $V_T=11.83\pm0.09$ (\citeauthor{RC3}) and $V-R = 0.48$, which is 
the median of four measurements by \cite{SV96}. This yields
$R_T\simeq  11.35\pm0.09$ for NGC~4606, hence the difference in $R$-band
total magnitudes is $3.13\pm 0.10$. For Gaussian-distributed errors, this
leads to a 90\%
probability that NGC~4606 is too faint to destroy the isolation of the M60
CG.
The errors here do not include uncertainties in the colors nor in the
calibration.

Hence, 
\emph{the M60 compact group has a good probability of being isolated}
(according to Hickson's criterion),
but one cannot statistically rule out that NGC~4606 is bright enough in
the $R$ band to spoil the group's isolation.

\section{SBF distance data}
\label{data}
As part of the ACS Virgo Cluster Survey \citep{Cote+04},
\cite{Mei+07} analyzed
Hubble 
Space Telescope images to measure distances to 84 
Virgo cluster ellipticals and S0s using the  SBF method.

\citeauthor{Mei+07} calibrated the SBF distances by fitting the trend of SBF
apparent magnitude vs. $(g_{475}-z_{850})_0$ color. They provided SBF
distances using three 
 different fits: \emph{linear}, \emph{polynomial}, and
\emph{broken-linear}. 
\citeauthor{Mei+07} expressed their preference for
the (4-parameter) broken-linear calibrated SBF distances. They noted that the $\chi^2$ of
their broken-line and (4-parameter) 4th-order 
polynomial fits were equally good, while
their linear (2 parameter) fit produced a slightly greater $\chi^2$.
They also remarked that the
broken-line fit had a smaller $\chi^2$ than the polynomial fit if 
the (three) galaxies redder than 
$(g_{475}-z_{850})_0 = 1.5$ were excluded.

According to Table 2 of
\citeauthor{Mei+07},  M60 turns out to be the reddest  (and 3rd
brightest) galaxy  in Virgo, with 
$(g_{475}-z_{850})_0=1.56$.
So, one infers that the polynomial calibration is superior for M60,
while the broken-line calibration is 
better for the three other galaxies of the
M60 CG.
I thus also consider a \emph{mixed} calibration which is broken-linear for 
$(g_{475}-z_{850})_0<1.5$ and polynomial for $(g_{475}-z_{850})_0\geq1.5$
(the broken-linear and polynomial fits intersect at this critical color, so
the mixed calibration is continuous).

\section{Analysis}
\label{anal}

\begin{figure}[ht]
\centering
\includegraphics[width=0.9\hsize]{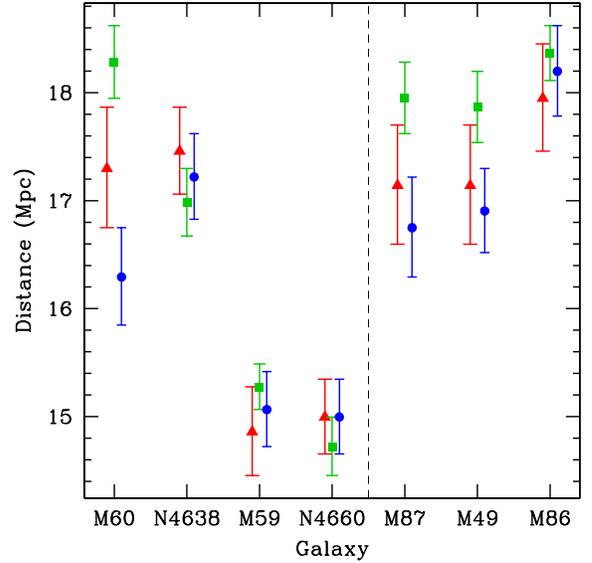}
\caption{Surface brightness fluctuation distances (from Mei et al. 2007) for
  the 4 ellipticals in the M60 compact group 
(\emph{left}) and for the three other
  brightest ellipticals in the cluster  (\emph{right}).
The \emph{red triangles}, \emph{green squares} and \emph{blue circles}
represent the distance measurements using the broken-line, linear, and
polynomial calibrations, respectively.}
\label{distplot}
\end{figure}

Table~\ref{cgtab} shows the data for the 4 group members for which SBF distance
measurements are available.
\begin{table*}[ht]
\begin{center}
\caption{Data including SBF distance moduli to the galaxies in the M60 compact
  group\label{cgtab} 
}
\begin{tabular}{cccccccccccc}
\hline
\hline
\multicolumn{3}{c}{Galaxy} & & RA & Dec &  type & $B_T$ & $v$ &
\multicolumn{3}{c}{distance modulus} \\
\cline{1-3}
\cline{5-6}
\cline{10-12}
Messier & NGC & VCC & &
\multicolumn{2}{c}{(J2000)} & & & ($\rm km \, s^{-1}$)
&  broken-line & linear & polynomial \\
\hline
59 & 4621 & 1903 & & $\rm 12^h42^m02\fs3$ & $+11^\circ38'49''$ & E5 &
10.57 & 
\ \ 410 &
$30.86\pm0.06$  & $30.92\pm0.03$  & $30.89\pm0.05$  \\ 
--- & 4638 & 1938 & & $\rm 12^h42^m47\fs4$ & $+11^\circ26'33''$ & S0 &
12.13 & 1164 & 
$31.21\pm0.05$  & $31.15\pm0.04$  & $31.18\pm0.05$  \\
---  & 4647 & 1972 & & $\rm 12^h43^m32\fs3$ &
$+11^\circ34'55''$ & Sc & 11.94 & 
1422 & --- & --- & --- \\
60 & 4649 & 1978 & & $\rm 12^h43^m40\fs0$ & $+11^\circ33'09''$ & E2 &
\ \ 9.81 & 1117 & 
$31.19\pm0.07$  & $31.31\pm0.04$  & $31.06\pm0.06$  \\
--- & 4660 & 2000 & & $\rm 12^h44^m32\fs0$ & $+11^\circ11'26''$ & E5 &
12.16 
& 1083 & 
$30.88\pm0.05$  & $30.84\pm0.04$  & $30.88\pm0.05$  \\ 
\hline
\end{tabular}
\end{center}
Notes: positions, types and heliocentric 
velocities are from NED, magnitudes from RC3, and
distance moduli from \cite{Mei+07}. 
\end{table*}
Figure~\ref{distplot} illustrates the distances to the 4 ellipticals in the M60
CG and to the three brightest Virgo galaxies (besides M60): \object{M87},
\object{M49} and \object{M86}.
M60 and NGC~4638 appear to be located at roughly the same distance as the
three luminous Virgo galaxies, M87, M49 and M86. On the other hand, 
M59 and NGC~4660, whose
SBF distances are consistent (regardless of the calibration used)
appear to lie roughly
2 Mpc closer to us.

Assuming Gaussian errors in the distance moduli, the distribution of 
the difference in distances of galaxies 1 and 2 with measured distance moduli
$\mu_1$ and $\mu_2$ and uncertainties $\sigma_1$ and $\sigma_2$ is a Gaussian
with mean $\mu_2-\mu_1$ and distribution $\sqrt{\sigma_1^2+\sigma_2^2}$.
Hence, the probability that the difference in distance moduli of the two
galaxies is greater than $\Delta \mu$ is
\begin{equation}
P(\Delta\mu) = {1\over2}\,\left \{1 + {\rm erf} \left [{\Delta \mu - (\mu_2-\mu_1) \over
  \sqrt{2\,(\sigma_1^2+\sigma_2^2)}} \right ]\right \} \ ,
\label{Pofdmu}
\end{equation}
where ${\rm erf}(x)$ is the error function.
Expressing the distance difference $\Delta D$ in terms of the difference in
distance moduli $\Delta\mu$ as
$\Delta D = 2\,D(\overline\mu) \,\sinh (0.1\,\ln 10\,\Delta \mu )$ and using
Eq.~(\ref{Pofdmu}), 
the minimum difference in distances 
of two galaxies is
\begin{eqnarray}
(\Delta D)_{\rm min} &=& 2\,D(\overline \mu) \nonumber
\\
&\mbox{}& \times
\sinh \left \{{\ln 10\over 10}\,\left [2\,\sigma_{\rm rms} \,{\rm erf}^{-1}
  (2P\!-\!1) + \Delta\mu \right ] \right \}
\ ,
\label{dmin}
\end{eqnarray}
where $y = {\rm erf}^{-1} (x)$ is the inverse error function, i.e. ${\rm
  erf}(y) = x$.

Table~\ref{minlos} provides the minimum distance difference between various
pairs of 
galaxies of the M60 CG, using Eq.~(\ref{dmin}). 
\begin{table}[ht]
\caption{Minimum line-of-sight separation of galaxy pairs\label{minlos}}
\begin{center}
\tabcolsep 2pt
\begin{tabular}{ccccr@{\ \ \ \ \ }}
\hline
\hline
Galaxy 1 & Galaxy 2 & SBF calibration & \multicolumn{2}{c}{$(D_2-D_1)_{\rm
    min} /(1\,\rm Mpc)$} \\  
\cline{4-5}
& & & $P=0.95$ & \multicolumn{1}{r}{$P=0.99$} \\
\hline
M59 & M60 & broken-linear & 1.32 & 0.85 \\
M59 & M60 & polynomial & 0.30 & --0.08 \\
M59 & M60 & linear & 2.37 & 2.11 \\
\bf M59 & \bf M60 & \bf mixed & \bf 0.43 & \bf 0.02 \\
\hline
\bf M59 & \bf NGC~4638 & \bf broken-linear & \bf 1.64 & \bf 1.25 \\
M59 & NGC~4638 & polynomial & 1.29 & 0.93 \\
M59 & NGC~4638 & linear & 1.10 & 0.84 \\
\hline
NGC~4660 & M60 & broken-linear & 1.25 & 0.82 \\
NGC~4660 & M60 & polynomial & 0.37 & --0.01 \\
NGC~4660 & M60 & linear & 2.85 & 2.56 \\
\bf NGC~4660 & \bf M60 & \bf mixed & \bf 0.37 & \bf --0.01 \\
\hline
\bf NGC~4660 & \bf NGC~4638 & \bf broken-linear & \bf 1.59 & \bf 1.23 \\
NGC~4660 & NGC~4638 & polynomial & 1.36 & 1.00 \\
NGC~4660 & NGC~4638 & linear & 1.58 & 1.30 \\
\hline
M59+NGC~4660 & M60 & broken-linear & 1.39 & 0.99 \\
M59+NGC~4660 & M60 & polynomial & 0.44 & 0.09 \\
M59+NGC~4660 & M60 & linear & 2.69 & 2.44 \\
\bf M59+NGC~4660 & \bf M60 & \bf mixed & \bf 0.52 & \bf 0.17 \\
\hline
\end{tabular}
\end{center}
Notes: the lines with \emph{mixed} SBF calibrations use 
broken-line and polynomial SBF calibrations for Galaxy 1 
and M60, respectively, which appear to be the most appropriate
calibrations for those galaxies.  For each combination of galaxies, the most
suitable 
set of SBF calibrations are highlighted in bold.
\end{table}
All three 
SBF estimators indicate that 
\emph{NGC~4638 is at least 800 kpc more distant than M59 and 1 Mpc more distant than NGC~4660} 
(both at the 99\% confidence
level).
Moreover, using the broken-linear or linear SBF calibrations, M60
must lie at
least 0.82 Mpc (99\% confidence) further away than either M59 or NGC~4660.
On the other hand, the SBF distances determined with the polynomial or mixed
  calibrations produce
consistent distances 
between M60 and either M59 or NGC~4660.
However, one can combine the distances to M59 and NGC~4660 to obtain a $\sqrt{2}$ smaller
uncertainty in the distance of that galaxy pair.
M60 then turns out to be
440 or 520 kpc further away than the pair
(at the 95\%
confidence level), depending on which of the polynomial or broken-linear SBF
  calibrations is used to estimate the distance of the pair.

\section{Discussion}
\label{discus}
What is the maximum line-of-sight
separation that is allowed for a
galaxy pair located within a dense group of galaxies?
Or
equivalently, what is the maximum line-of-sight size of a dense group of
galaxies? 

One can specify that the maximum line-of-sight
separation between galaxy pairs in a dense group must be smaller than twice
its projected diameter or, alternatively, twice the 84th percentile
(corresponding to $+1\,\sigma$ for a Gaussian distribution)
of the
projected diameters of HCGs.
Given that the angular radius of the Hickson circle
of the M60 CG is $0\fdg38$ \citep{Mamon89},\footnote{This angular radius is not
  affected by the removal of 
NGC~4647 (see Fig.~\ref{isol}).} and given its (error-weighted) mean
(mixed SBF calibration) distance of 15.94 Mpc (see
Fig.~\ref{distplot}), the projected 
radius of the Hickson circle is 106 kpc.
In comparison, the median projected radius of the 68 HCGs with at least 4
accordant 
velocities\footnote{I exclude
\object{HCG 54}, the HCG with the smallest 
projected radius, because it does not constitute a group of galaxies, 
appearing instead
to be either a group of 
H\,{\sc \large{ii}}
regions in a single galaxy \citep{Arkhipova+81} or the end result of the
merger of two 
disk galaxies \citep{VerdesMontenegro+02}.} is 56 kpc, and 106 kpc
corresponds to the 87th percentile of the distribution of HCG projected
radii.
Both 
criteria  are therefore virtually identical.
I thus adopt a maximum line-of-sight size of $4\times 106 = 424$
kpc.\footnote{This is a very liberal choice: 
in a recent analysis of compact groups of galaxies in the
  Millennium survey, \cite{DRMM08} show that maximum three-dimensional
  separations of $160 \, h^{-1} \, \rm kpc \simeq 224\,\rm kpc$ are required
  to produce physically dense groups of
  galaxies, selected with Hickson's criteria, whose line-of-sight sizes are
  on average equal to their projected diameters. Adopting a smaller maximum
  line-of-sight dimension will reinforce the conclusion that the M60 CG is
  caused by a chance alignment of galaxies.}

According to Table~\ref{minlos},
 \emph{it is highly unlikely that NGC~4638 is part of a dense group or pair
   containing M59 and    NGC~4660}.
Still, 4 galaxies remain once  NGC~4638 is omitted.
Nevertheless, \emph{M60 cannot be part of the dense group or pair containing
  M59 and NGC~4660}, at a 99\% (broken-linear or linear SBF
calibrations) or 95\% (polynomial SBF calibration) confidence level.
Therefore, one can state with high confidence that 
among the four early-type galaxies in the M60 CG, \emph{M59 and
  NGC~4660 cannot constitute a dense group of galaxies with M60 and NGC~4638.}

The M60 CG in Virgo is just one example of a Hickson-like compact group.
Up to now, SBF distances have been measured for galaxies out to $v=4000 \, \rm km
\, s^{-1}$ \citep{Tonry+01} and the ACS Virgo Cluster Survey has measured
distances to galaxies as faint as
$B_T = 
16$ \citep{Mei+07}. 
There is one HCG within these distance and magnitude limits: \object{HCG~68}
($v=2400 \, \rm km \, s^{-1}$, \citealp{HMdOHP92}, 2 ellipticals and 2 S0s, all
with $B_T\leq 14.56$, \citealp{HKA89}, plus one Sbc), whose nature could
therefore be probed in the same way as for the M60 CG.
In the near future, SBF distances should become available for fainter and
more distant early-type galaxies, allowing for direct line-of-sight
analyses, similar to the present one, for additional HCGs.

Alternatively, 
the nature of the
the full set of HCGs can be assessed by 
confronting the
observational properties of these exceptionally dense galaxy systems with
those constructed 
using either cosmological hydrodynamical simulations
that can resolve sufficiently small galaxies, or alternatively, galaxy formation
simulations based upon realistic galaxy positions obtained from
high-resolution cosmological dark matter simulations.
Using this second approach, \cite{McCEP08} and \cite{DRMM08} have recently shown that
roughly 
half of the
Hickson-like CGs with at least 4 accordant
velocities 
are chance
alignments of galaxies, the precise fraction depending on the cut-off in
maximum line-of-sight size, and on the galaxy formation code ran on the outputs of
the Millennium dark matter simulations \citep{Springel+05}.

\begin{acknowledgements}
The author thanks Simona Mei for several very useful discussions on the SBF
distance indicator, Sandy Faber on SDSS photometry of very bright galaxies,
and the anonymous referee for useful comments.

This research has made use of the NASA/IPAC Extragalactic Database (NED)
which is operated by the Jet Propulsion Laboratory, California Institute of
Technology, under contract with the National Aeronautics and Space
Administration, the HyperLEDA database \citep{Paturel+03}, the SDSS
and GoogleSky
  ({\tt 
http://earth.google.com}). 

Funding for the Sloan Digital Sky Survey (SDSS) and SDSS-II has been provided
by the Alfred P. Sloan Foundation, the Participating Institutions, the
National Science Foundation, the U.S. Department of Energy, the National
Aeronautics and Space Administration, the Japanese Monbukagakusho, and the
Max Planck Society, and the Higher Education Funding Council for England. The
SDSS Web site is http://www.sdss.org/.
The SDSS is managed by the Astrophysical Research Consortium (ARC) for the
Participating Institutions. The Participating Institutions are the American
Museum of Natural History, Astrophysical Institute Potsdam, University of
Basel, University of Cambridge, Case Western Reserve University, The
University of Chicago, Drexel University, Fermilab, the Institute for
Advanced Study, the Japan Participation Group, The Johns Hopkins University,
the Joint Institute for Nuclear Astrophysics, the Kavli Institute for
Particle Astrophysics and Cosmology, the Korean Scientist Group, the Chinese
Academy of Sciences (LAMOST), Los Alamos National Laboratory, the
Max-Planck-Institute for Astronomy (MPIA), the Max-Planck-Institute for
Astrophysics (MPA), New Mexico State University, Ohio State University,
University of Pittsburgh, University of Portsmouth, Princeton University, the
United States Naval Observatory, and the University of Washington.

\end{acknowledgements}

\bibliography{master}
\end{document}